\documentclass[%
reprint,
 amsmath,amssymb,
 aps,
 prl,
]{revtex4-1}

\usepackage{graphicx}
\usepackage{dcolumn}
\usepackage{bm}

\usepackage{color} 
\usepackage{tabularx}

\begin{document}

\preprint{APS/123-QED}

\title{In situ observation of Hall Magnetohydrodynamic Cascade in Space Plasma}

\author{Riddhi Bandyopadhyay}
\affiliation{Department of Physics and Astronomy, University of Delaware, Newark, Delaware 19716, USA}

\author{Luca Sorriso-Valvo}
\affiliation{Departamento de F\'isica, Escuela Polit\'ecnica Nacional, 170517 Quito, Ecuador}
\affiliation{Istituto per la Scienza e Tecnologia dei Plasmi, Consiglio Nazionale delle Ricerche, 87036 Bari, Italy}

\author{Alexandros Chasapis}
\affiliation{Laboratory for Atmospheric and Space Physics, University of Colorado Boulder,  Boulder, Colorado, USA}	

\author{Petr Hellinger}
\affiliation{Astronomical Institute, AS CR, Prague, Czech Republic}
\affiliation{Institute of Atmospheric Physics, AS CR, Prague, Czech Republic}

\author{William~H. Matthaeus}
\affiliation{Department of Physics and Astronomy, University of Delaware, Newark, Delaware 19716, USA}
\affiliation{Bartol Research Institute, University of Delaware, Newark, Delaware 19716, USA}	

\author{Andrea Verdini}
\affiliation{Dipartimento di Fisica e Astronomia, Universit\'a degli Studi di Firenze, 50125 Firenze, Italy}
\affiliation{INAF, Osservatorio Astrofisico di Arcetri, Largo E. Fermi 5, I-50125 Firenze, Italy}

\author{Simone Landi} 
\affiliation{Dipartimento di Fisica e Astronomia, Universit\'a degli Studi di Firenze, 50125 Firenze, Italy}
\affiliation{INAF, Osservatorio Astrofisico di Arcetri, Largo E. Fermi 5, I-50125 Firenze, Italy}

\author{Luca Franci} 
\affiliation{School of Physics and Astronomy, Queen Mary University of London, London, UK}
\affiliation{INAF, Osservatorio Astrofisico di Arcetri, Largo E. Fermi 5, I-50125 Firenze, Italy}

\author{Lorenzo Matteini}
\affiliation{LESIA, Observatoire de Paris, Meudon, France}
\affiliation{INAF, Osservatorio Astrofisico di Arcetri, Largo E. Fermi 5, I-50125 Firenze, Italy}

\author{Barbara~L. Giles}
\affiliation{NASA Goddard Space Flight Center, Greenbelt, Maryland 20771, USA}

\author{Daniel~J. Gershman}
\affiliation{NASA Goddard Space Flight Center, Greenbelt, Maryland 20771, USA}

\author{Craig~J. Pollock}
\affiliation{Denali Scientific, Fairbanks, Alaska 99709, USA}

\author{Christopher~T. Russell}
\affiliation{University of California, Los Angeles, California 90095-1567, USA}

\author{Robert~J. Strangeway}
\affiliation{University of California, Los Angeles, California 90095-1567, USA}

\author{Roy~B. Torbert}
\affiliation{University of New Hampshire, Durham, New Hampshire 03824, USA}

\author{Thomas~E. Moore} 
\affiliation{NASA Goddard Space Flight Center, Greenbelt, Maryland 20771, USA}
	
\author{James~L. Burch}
\affiliation{Southwest Research Institute, San Antonio, Texas 78238-5166, USA}

\date{\today}

\begin{abstract}
We present estimates of the turbulent energy cascade rate, derived from a { Hall-Magnetohydrodynamic (MHD)
third-order law}. We compute the contr  ibution from the Hall term and the MHD term to the energy flux. {Magnetospheric MultiScale (MMS)} data, accumulated in the magnetosheath and the solar wind, are compared with previously established simulation results. Consistent with the simulations, we find that at large
(MHD) scales the MMS observations exhibit a clear inertial range, dominated by the MHD flux. In the sub-ion
range the cascade continues at a diminished level via the Hall term, and the change becomes more pronounced as the plasma beta increases. Additionally, the MHD contribution to interscale energy transfer remains important at smaller scales than previously thought. Possible reasons are offered for this unanticipated result.
\end{abstract}

\maketitle

Fully developed turbulence is characterized by scale-invariant energy transfer across the inertial range of length scales, where nonlinear terms dominate the dynamics \cite{Kolmogorov1941c}. 
In the solar wind, planetary magnetospheres, and other turbulent astrophysical plasmas, large scale energy,
in the form of velocity shears and electromagnetic fluctuations, is transferred across scales and dissipated at small scales. This process is known as \textit{energy cascade}. Turbulent energy cascade and dissipation have important effects in space
and astrophysical systems and are considered an important source for the observed plasma heating \cite{Richardson1995GRL} and acceleration of energetic particles.

In homogeneous, neutral fluid turbulence, 
the Kolmorov-Yaglom law \cite{Karman1938PRSL,Kolmogorov1941a} quantifies the mean energy dissipation rate in terms of longitudinal, third-order structure functions. The Kolmogorov-Yaglom third-order law is extended to the case of plasmas, in the incompressible magnetohydrodynamic (MHD) description, by Politano and Pouquet \cite{Politano1998GRL,Politano1998PRE,Sorriso-Valvo2002PoP}. This MHD theory accounts for
the \emph{incompressive channel} of the inertial range 
energy cascade. For plasmas with small density fluctuations, 
such as the cases presented here, the incompressive transfer is expected to contribute 
a majority of the total energy transfer \cite{Yang2017PoF,Hadid2018PRL,Montagud-Camps2018ApJ}. 

In the rapid streaming of the solar wind (at mean
speed $\langle \mathbf{V} \rangle$),
the Taylor hypothesis \cite{Taylor1938PRSLA} 
($\mathbf{r} = t\langle \mathbf{V} \rangle$) 
permits conversion of space ($\mathbf{r}$) and time ($t$) arguments.
Then the 
Politano-Pouquet law prescribes the linear scaling of the mixed, third-order moment { 
\begin{equation}
Y(\ell) \equiv \langle 
\delta v_{l}(|\delta \mathbf{v}|^2+|\delta \mathbf{b}|^2)-2\delta b_{l}(\delta \mathbf{v}\cdot\delta \mathbf{b}) \rangle = -\frac{4}{3}  \varepsilon \ell
\label{Y}
\end{equation}
where $\delta$ indicates
an increment, e.g., 
$\delta\psi(t,\delta t)=\psi(t+\delta t)-\psi(t)$ 
for a generic field $\psi$ and $\ell =
\delta t \langle V \rangle$.
In MHD, we compute
increments 
of the plasma velocity $\mathbf{v}$ 
or the magnetic field, $\mathbf{b}=\mathbf{B}/\sqrt{4\pi\rho}$
(in Alfv\'en units, mass density $\rho$) 
using a temporal scale $\delta t$.} 
The subscript $l$ indicates longitudinal components, and $\varepsilon$ is the mean energy transfer rate. 

{ Assuming that the turbulence 
is approximately statistically stationary and 
homogeneous \cite{Matthaeus1982bJGR}, Eq. (\ref{Y})
enables estimation of the fluid-scale energy transfer rate.}
The Politano-Pouquet law, in its isotropic form, has been verified in solar wind studies \cite{MacBride2005SW,Sorriso-Valvo2007PRL,Coburn2012ApJ,Banerjee2016ApJL}, and more recently in the terrestrial magnetosheath \cite{Hadid2018PRL,Bandyopadhyay2018bApJ} and magnetospheric boundary layer \cite{Sorriso-Valvo2019PRL}. 
The cascade rate measured this way 
is shown to sufficiently account for the solar wind heating \cite{Vasquez2007JGR,MacBride2008ApJ,Marino2008ApJ}. 
The presence of a 
significant mean DC magnetic field in the solar wind 
leads to an expectation 
of spectral anisotropy \cite{Shebalin1983JPP}. 
However, even when the anisotropic form of the Politano-Pouquet law is used to derive the solar wind heating rate, 
the results are fairly close to that obtained from the isotropic  scaling law \cite{Osman2011PRL,Verdini2015ApJ}.

The single-fluid
MHD phenomenology is only 
suitable in the fluid regime -- 
large length scales and low-frequencies. 
As smaller length scales are approached, 
near the ion gyro-radius $(\rho_{\mathrm{i}})$ or ion-inertial length $(d_{\mathrm{i}})$, the nature of the 
cascade is expected to change. 
For example, ideal motions of the 
plasma should render the magnetic field ``frozen in" 
the electron fluid motions at velocity $ \mathbf{v}_e $,
rather than frozen into the MHD fluid frame at 
(proton) velocity $\mathbf{v}$. 
Near the kinetic scales, 
to first-order approximation, kinetic physics can be partially included via the Hall electric field 
in the fluid model \cite{Papini2019ApJ}. 
Accordingly,
employing 
the equations of incompressible Hall MHD, a scaling law for the third-order structure functions, 
analogous to its MHD counterpart, can be derived to obtain the energy cascade flux at 
the scale of interest \cite{Galtier2008PRE,Hellinger2018ApJL,Ferrand2019}. In the Hall MHD formulation, the third-order moment scaling law includes the additional Hall term { 
\begin{equation}
H(\ell) \equiv \langle 2 \delta  b_l(\delta \mathbf{b}\cdot\delta \mathbf{j})-\delta j_l|\delta \mathbf{b}|^2\rangle.
\label{H}
\end{equation} }
{ Here, $\mathbf{j}$ is the electric current density in Alfv\'en units: $\mathbf{j} = \mathbf{v}-\mathbf{v}_{\mathrm{e}}$; where, $\mathbf{v}$ is the proton velocity and $\mathbf{v}_{\mathrm{e}}$ is the electron velocity.
When the displacement current is 
neglected, this is equivalent to $\mathbf{j} = \nabla \times
\mathbf{b}$.} Hellinger \textit{et al.} \cite{Hellinger2018ApJL} derive the Hall contribution to $Y$  as $H$, neglecting an additional 
contribution equal to $-H/2$ \cite{Ferrand2019},
so that complete scaling law reads: 
\begin{equation}
Y + \frac{1}{2}H=-\frac{4}{3}  \varepsilon \ell
\label{hall}.
\end{equation}

{ 
In applying this equation to spacecraft observations in a weakly-collisional plasma, as we do here, or to simulations as in \cite{Hellinger2018ApJL}, we recognize that such a formulation is incomplete and lacks many kinetic effects. However, in general one expects that such additional effects will be additive, and therefore 
from a theoretical perspective
the Hall third-order law represents a better measure of the energy-transfer rate near the kinetic scales, compared to MHD.}
The linear scaling, Eq. (\ref{hall}), has been recently applied to 
hybrid-kinetic numerical
simulations \cite{Hellinger2018ApJL,Ferrand2019}, where the Hall-MHD-generalized flux becomes
dominant at small scales, continuing a scale-invariant
cascade further into the sub-proton range. However,
the energy-cascade flux decreases near the kinetic scale,
even after including the contribution from the Hall-term. This decrease 
is stronger in high-$\beta$ plasma.

In this work, we study the Hall-MHD third-order law using in-situ data from the Magnetospheric Multiscale (MMS) spacecraft~\cite{Burch2016SSR, Pollock2016SSR, Russell2016SSR} and compare the results directly with the analysis of two-dimensional hybrid-kinetic numerical simulations \cite{Hellinger2018ApJL}. The MMS results exhibit similarities to
and, perhaps surprisingly, some contrasts with, the simulation results and baseline theoretical expectations.  

\begin{table*}
\caption{ Some Plasma Parameters of the Selected Intervals. SW $\equiv$ Solar Wind, MSH $\equiv$ Magnetosheath.}
	\label{tab:overview}
		\begin{tabularx}{\textwidth}{X X X X X X X X X X}
			\hline \hline
			Interval &
			$\beta_{\mathrm{i}}$ &
			$|\langle \mathbf{B} \rangle|$ &
			$B_{\mathrm{rms}} /|\langle \mathbf{B} \rangle|$ &
			$\rho_{\mathrm{rms}} /\langle \rho \rangle$ &
			$M_{\mathrm{t}}$ &
			$|\langle \mathbf{V} \rangle|$ &
			$V_{\mathrm{A}}$
			& $d_{\mathrm{i}}$ & $L_{\mathrm{corr}}$
			\\
			 & & $(\mathrm{nT})$ &  &  & & $(\mathrm{km~s^{-1}})$ & $(\mathrm{km~s^{-1}})$ &  $(\mathrm{km})$ & $(\mathrm{km})$\\
			\hline
			 SW & 0.4 \footnote{Temperature provided by Wind~\cite{Ogilvie1995SSR, Lepping1995SSR} spacecraft is used to compute the beta value and the turbulent Mach number in the solar wind.\label{foot:temp}} & 7.4 & 0.3 & 0.08 & 0.3\textsuperscript{\ref{foot:temp}} & 330 & 51 & 75 & $11 \times 10^{4}$ \\
			 MSH & 17 & 3.4 & 2.5 & 0.2 & 0.2 & 278 & 18 & 56 & 1350 \\
			\hline \hline
		\end{tabularx}
\end{table*}

To study the Hall MHD third-order law, we use burst resolution MMS data accumulated in two distinct turbulent plasma environments. The first one is an hour-long solar wind (SW) interval on 2017-11-26 from 21:09:03 to 22:09:03 UTC, far from the Earth's bow shock. We do not find any signature of reflected ions from the bow shock, so the solar wind interval can be considered to be ``pristine.'' 

{ The second dataset that we use is an MMS interval sampled in the terrestrial magnetosheath (MSH) on 2018-10-27 from 09:13:13 to 09:57:43 UTC. Here, the plasma beta is much higher with an average value of $\beta_{\mathrm{i}} = 17$.} Table \ref{tab:overview} reports some important plasma parameters for the two selected intervals. From table \ref{tab:overview}, we note that for the chosen intervals, the flow speed is larger than the Alfv\'en speed, indicating that Taylor hypothesis is expected to be valid. 
{ These MMS intervals are typical -- analyses of other solar wind and magnetosheath samples (not shown, but see Supplementary Material) exhibit similar properties and conclusions~\citep[also see][]{Huang2014ApJL,Huang2017ApJL,Hadid2018PRL}.}

\begin{figure}
	\begin{center}
		\includegraphics[scale=1]{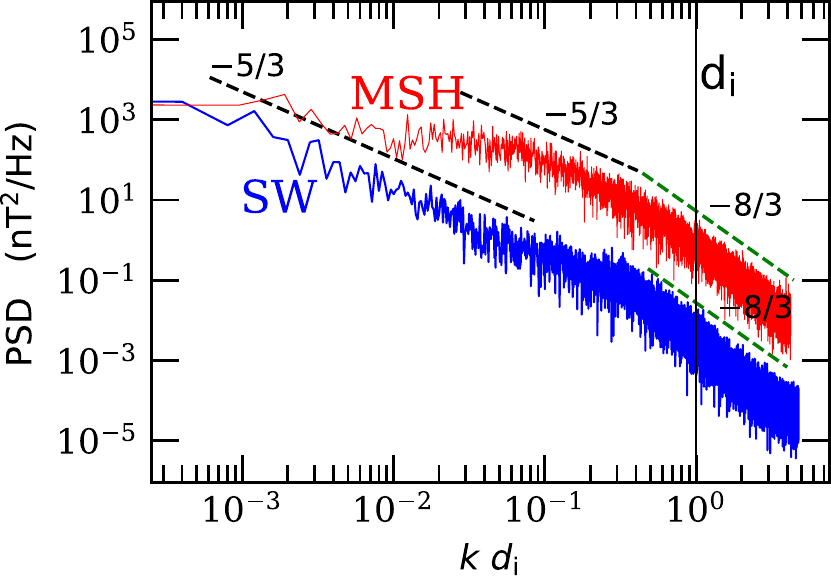}
		\caption{Magnetic field spectra for, solar wind (SW, in blue), and Magnetosheath (MSH, in red) interval.  The solid vertical line represents $k d_{\mathrm{i}} = 1$ with the wave vector $k \simeq (2\pi f)/|\langle \mathbf{V} \rangle|$.}
		\label{fig:spec}
	\end{center}
\end{figure}

Figure~\ref{fig:spec} illustrates the power spectral density (PSD) of the magnetic field fluctuations for the two chosen intervals, plotted against $k d_{\mathrm{i}}$. The level of fluctuations is considerably higher in the magnetosheath interval than in the solar wind. Both spectra exhibit Kolmogorov-like
$-5/3$ scaling in the inertial range, followed by a steepening near $k d_{\mathrm{i}} = 1$. However, the solar-wind spectrum has a significantly broader bandwidth of inertial range, 
representative of 
a larger and 
higher Reynolds number system,
compared to the magnetosheath interval \cite{Huang2014ApJL,Huang2017ApJL}.

Having shown that the two chosen intervals exhibit extended, inertial-range Kolmorogov spectra, we compute the energy flux from Eq. (\ref{hall}). The required 
electron and proton moments provided by FPI in the solar wind are processed using the method described in \cite{Bandyopadhyay2018aApJ} to exclude instrumental artifacts that arise when FPI operates in the solar wind. All analyses are performed over the four MMS spacecraft and then averaged. In the solar wind, a spintone signal is introduced likely due to issues in the very low-energy response of the electrons, and a systematic offset in the ion and electron velocities. Therefore, we use the curlometer-based~\cite{Dunlop1988ASR} current for the Hall term.

Figure~\ref{fig:eps} shows the scaling of the third-order structure functions, decomposed into the MHD $(-Y)$ and Hall MHD $(-H/2)$ terms, from Eq. (\ref{hall}), with spatial lag in units of $d_{\mathrm{i}}$. Only parts of the structure functions that lie well above the 
instrumental noise level are plotted here. A roughly linear scaling is observed in the inertial range for both samples. In both samples, the MHD component $-Y$ shows better scaling in the inertial range, where it is dominant with respect to the Hall term $-H/2$. The latter has more defined scaling at scales near or below $d_{\mathrm{i}}$, where its contribution to the energy transfer becomes of the same order as for the MHD term. Note that although the structure functions exhibit large fluctuations at scales near and larger than the correlation scale, within the narrow range demarcated by vertical dashed and solid lines the linear scaling is rather good and smooth (Fig.~\ref{fig:eps}, bottom panel).

\begin{figure}
	\begin{center}
		\includegraphics[width=\linewidth]{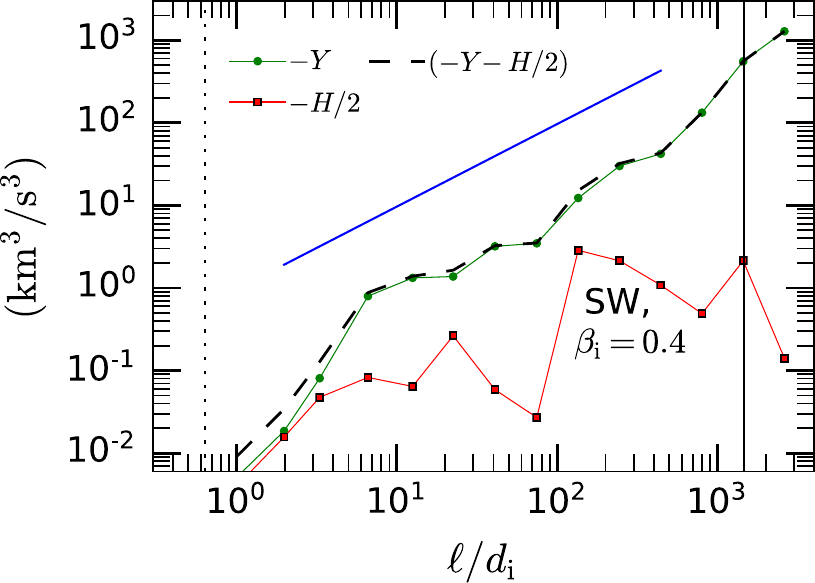}
		\includegraphics[width=\linewidth]{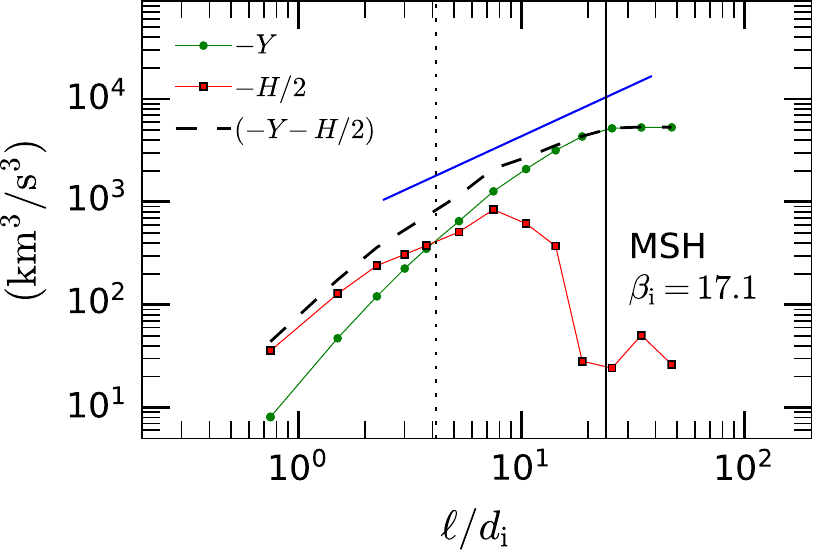}
		\caption{ MHD ($-Y$) and Hall ($-H/2$) structure function from generalized third-order law (Eq. \ref{hall}) from MMS data. Top: solar wind, $\beta_{\mathrm{i}} = 0.4$. Bottom: Magnetosheath, $\beta_{\mathrm{i}} = 17$. The proton gyro-radius, $\rho_{\mathrm{i}}$, is shown as a dotted, vertical line and the correlation length, $L_{\mathrm{corr}}$, is shown as a solid vertical line. A linear scaling with arbitrary offset is shown for reference.}
		\label{fig:eps}
	\end{center}
\end{figure}

\begin{figure}
	\begin{center}
		\includegraphics[scale=1]{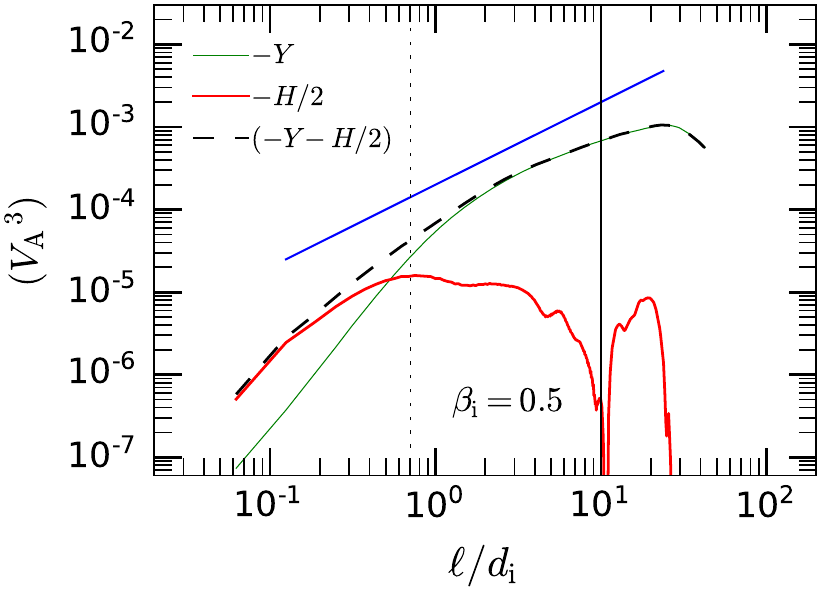}
		\includegraphics[scale=1]{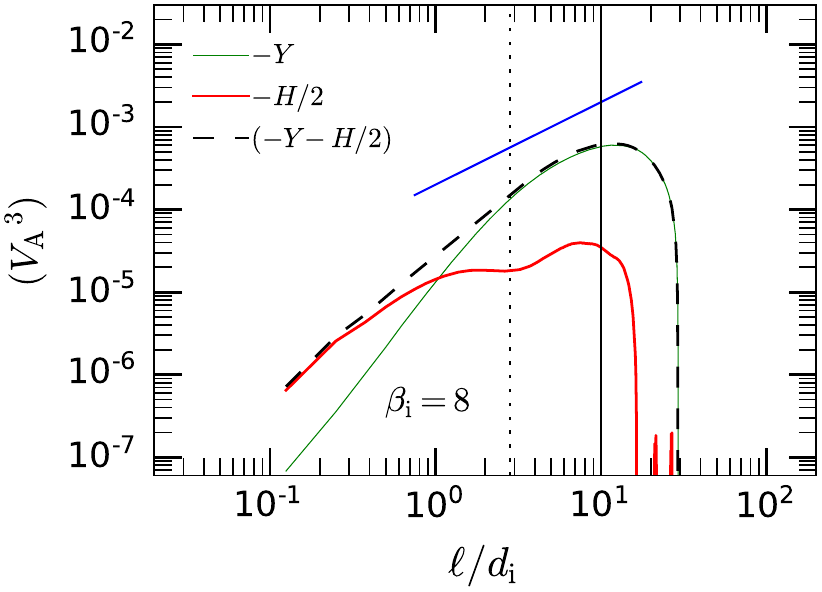}
		\caption{MHD ($-Y$) and Hall ($-H/2$) structure function from generalized third-order law (Eq. (\ref{hall})) from 2D hybrid-kinetic simulations. Top panel: $\beta_{\mathrm{i}}=0.5$. Bottom panel: $\beta_{\mathrm{i}}=8$. The dotted vertical line indicates the ion gyroradius, $\rho_{\mathrm{i}}$, and the solid vertical line indicates the correlation length $L_{\mathrm{corr}}$. A linear scaling with arbitrary offset is shown for reference.}
		\label{fig:sim}
	\end{center}
\end{figure}
In order to compare the above results with numerical simulations, we use two different two-dimensional runs  of the hybrid code CAMELIA~\citep{Franci2018JPCS}. The two runs have  $\beta_{\mathrm{i}}=0.5$ and $8$, chosen to probe variations of 
$\beta$ comparable to 
the contrast in the solar wind and magnetosheath plasma properties. The simulation box has the size $256 d_{\mathrm{i}} \times 256 d_{\mathrm{i}} $ for both runs. More detailed description of these runs can be found in \cite{Hellinger2018ApJL} and \cite{Franci2016ApJ}.

Figure~\ref{fig:sim} shows the Hall MHD third-order law~(\ref{hall}) \cite{Ferrand2019}, in Alfv\'en speed units, 
in a format similar to Fig.~\ref{fig:eps}. The transition between MHD and ion (or Hall) scales occurs roughly at the ion gyroradius $\rho_{\mathrm{i}}$ which is indicated by the dotted vertical line. The correlation length is about $10~d_{\mathrm{i}}$ for both runs and it is indicated by solid black vertical line.  The relatively small range of  linear scaling seen in the simulation results is typical of hybrid simulation~\cite{Parashar:ApJL2015} and, due to limits on available computing resources, is an unavoidable consequence of  limited scale and resolution.
However, there is reasonable level of
qualitative agreement with the observations. 
In particular, in the low-$\beta$ simulation the Hall term, $-H/2$, becomes relevant closer to the transition scale more prominently than in the high-$\beta$ case. 
Conversely the dominance of the MHD contribution 
is established more dramatically
at larger scales in the lower $\beta$ 
solar wind and lower beta simulation. 
Note that these 
include a corrected Hall-term contribution relative to earlier 
results \cite{Hellinger2018ApJL}. 

\begin{figure}
	\begin{center}
		\includegraphics[width=0.85\linewidth]{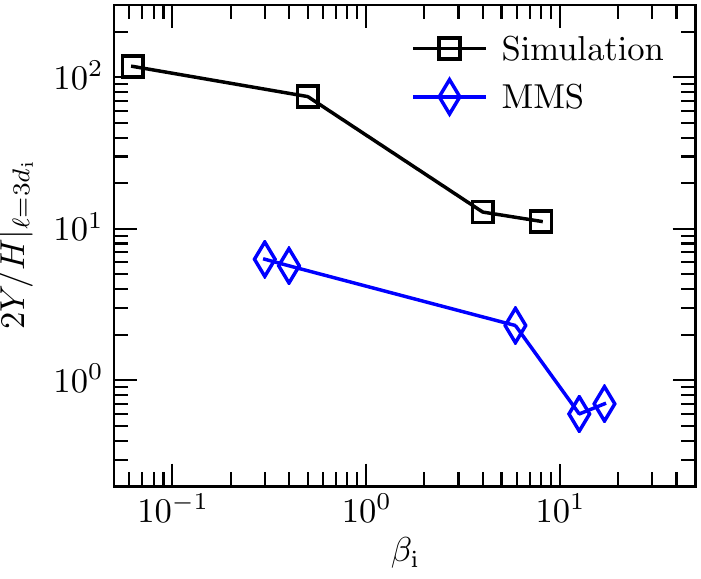}
		\caption{ Variation with ion beta
		of the ratio of the MHD 
		term $(-Y)$ to the Hall-MHD term 
		$(-H/2)$ in the third-order cascade rate at lag $\ell = 3\,d_{\mathrm{i}}$, 
		as obtained from  simulation and MMS data.}
		\label{fig:beta}
	\end{center}
\end{figure}

{ While the observational results behave qualitatively similar to the simulations near the kinetic scales and at larger scales, the comparison in the sub-proton range of scales
is less clear. Unlike in the numerical simulations, in the
observations, the Hall-contributed cascade becomes significant at much larger scale and the MHD-contributed cascade remains important at smaller scales.
Still, in all cases we can confirm that 
the Hall physics becomes increasingly important for the energy 
transfer at sub-proton scales. 
However, both simulation and in-situ observations have implicit limitations that may provide possible explanation
for the apparent (if somewhat subtle) 
differences.} 

{To support the conclusion that the Hall term becomes significant at larger scales in observations than in the simulations, we analyze a few more simulation and MMS dataset (see Supplementary material) with wide range of $\beta_{\mathrm{i}}$-values. We evaluate the ratio of the MHD to the Hall term in Eq.~\ref{hall} for a fixed length scale of $\ell = 3 d_{\mathrm{i}}$, and plot them with the $\beta_{\mathrm{i}}$-values in Fig.~\ref{fig:beta}. Although both simulations and observations exhibit a decreasing trend with increasing $\beta_{\mathrm{i}}$, the ratio is significantly smaller for the observational cases, compared to the simulations.}

There may be several reasons for this difference. The magnetosheath is a smaller system than the solar wind, and 
exhibits a significantly narrower range 
between kinetic scale (either $\rho_{\mathrm{i}}$ or $d_{\mathrm{i}}$) and the correlation scale. It is possible 
that the small separation of scales does not allow the two contributions to the energy flux to be sufficiently distinct \cite{Matthaeus2003GRL}. { Deviation from strict statistical homogeneity and incompressibility may also play a role.}  
A notable feature is that,
in both magnetosheath and solar wind cases, 
the Hall and standard-MHD cascade contributions 
(Fig. \ref{fig:eps})
become comparable at a few $d_i$, 
at nearly the scales where the corresponding 
kinetic range modifications~\citep{Matteini2016MNRAS} to the spectra begin to 
be seen (Fig. \ref{fig:spec}.) In the simulations,
the more dramatic crossover of Hall and MHD effects 
occurs at moderately 
smaller sub-$d_{\mathrm{i}}$ scales. 

At the same time, the hybrid-kinetic simulations are two-dimensional and admit rather low Reynolds number values; both of which may potentially alter the nature of energy cascade. Additionally, the hybrid simulations ignore the kinetic effects of electrons. With the current computational ability, three-dimensional hybrid simulations would be severely limited in Reynolds number, even more so in full kinetic simulations. { For these reasons, a precise quantitative correspondence between the observations and currently attainable simulations should not be expected.}

Finally, Table \ref{tab:eps} reports the approximate values of the inertial-range energy-transfer rate obtained for the two chosen intervals from the Hall MHD scaling law (Eq. (\ref{hall})). The second column denotes the total energy transfer rate from the Hall MHD law: $\epsilon_{\mathrm{inertial}} = -3 (Y + H/2) /4\ell$. The magnetosheath-energy decay rate is about three orders of magnitude larger than the interplanetary solar wind  \cite{Hadid2018PRL,Bandyopadhyay2018bApJ,Sorriso-Valvo2019PRL}. 
The final column is a rough estimate of the global energy decay rate, at the energy-containing scale, obtained from a von K\'arm\'an-Taylor \cite{Karman1938PRSL,Taylor1935,Hossain1995PoF,Wan2012JFM} phenomenology (see \cite{Bandyopadhyay2018bApJ} and the supplementary material). 
The von K\'arm\'an estimates are close to the inertial-scale ones from the third-order law. {Although the energy-cascade flux has a fairly constant value in the inertial range, all the MMS intervals, as well as the current and prior simulations~\citep{Hellinger2018ApJL} indicate that the flux on average decreases near the kinetic scales, even after including the Hall-contribution. This decrease is more prominent in the high-beta cases, and presumably is due to the onset of other kinetic effects at subproton scales~\citep{Yang2017PoP}.} Note that the value of proton beta does not affect the cascade rate, which is determined by the energy budget at the large-scales, as estimated by the von K\'arm\'an theory.

\begin{table}
\caption{Energy flux in units of $\mathrm{J~kg^{-1}~s^{-1}}$}
	\label{tab:eps}
		\begin{tabularx}{0.5\textwidth}{X X X}
			\hline \hline
			Interval &
			$\epsilon_{\mathrm{inertial}}$ &
			$\epsilon_{\mathrm{vK}}$
			\\
			\hline
			 SW  & $(0.9 \pm 0.1) \times 10^{3}$ & $0.7 \times 10^{3}$ \\
			 MSH  & $(3 \pm 0.5) \times 10^{6}$ & $2.6 \times 10^{6}$\\
			\hline \hline
		\end{tabularx}
\end{table}

Understanding how collisionless plasmas dissipate remains a topic of central importance in space physics, astrophysics, and laboratory plasma. In the recent years, it has become increasingly recognized that MHD description must be refined to clearly make connection with kinetic plasma dissipation. The present results provide a step towards understanding this problem. 
Based on the unprecedented capabilities of 
the MMS mission instrumentation, 
the findings
of this paper 
confirms the 
applicability of 
the Hall-modified
third-order order laws, 
as similar, but not identical behavior is seen in the transition to kinetic effects near proton scales. We note that a similar paper has recently been published~\citep{Andres2019PRL}, with results in the magnetosheath, while in the present work we also analyze solar wind results and compare them with simulations.
{  We may summarize 
by saying that 
the trend in relative strength of the Hall effect on the cascade
is confirmed in the observations and simulations, 
with respect to variation of beta and scale. However, the 
disparity in magnitude of both effects differs, 
presumably due to the effects described above. 
This raises a note 
of caution with regard to 
quantitative comparisons with observations,
due to limiting approximations of several types.} 
Clarification of these subtle differences awaits investigations with more advanced simulations and observational data, when available in the future.

\begin{acknowledgments} 
A.C. is supported by MMS/NASA under NNX14AC39G. L.S.V. is supported by the EPN Internal Project PII-DFIS-2019-01. W.H.M. is a member of the MMS Theory and Modeling team and is supported by NNX14AC39G and a NASA Supporting Research grant NNX14AC39G. P.H. acknowledges grant 18-08861S of the Czech Science Foundation. We are grateful to the MMS instrument teams for cooperation and collaboration in preparing the data. The data used in this analysis are Level 2 FIELDS and FPI data products, in cooperation with the instrument teams and in accordance with their guidelines. 
All MMS data are available at \url{ https://lasp.colorado.edu/mms/sdc/}.
\end{acknowledgments}

\section{Supplemental Material}

In this supplemental material, we present additional material which supplement the letter. In the first Secion, we show the time series of the two intervals discussed in detail in the main text, and normalize the derived cascade rates appropriately. In the seond Section, we present a set of 4 Hybrid-kinetic simulations with wide range of $\beta_{\mathrm{i}}$-values, to examine the effect of system size on the linear scaling of the third-order structure functions. In the last Section, we show the scaling of the MHD and Hall cascade rates for additional MMS magnetosheath intervals with broad $\beta_{\mathrm{i}}$-values.

\section{Overview}\label{sec:overview}
\begin{figure}
	\begin{center}
		\includegraphics[width=\linewidth]{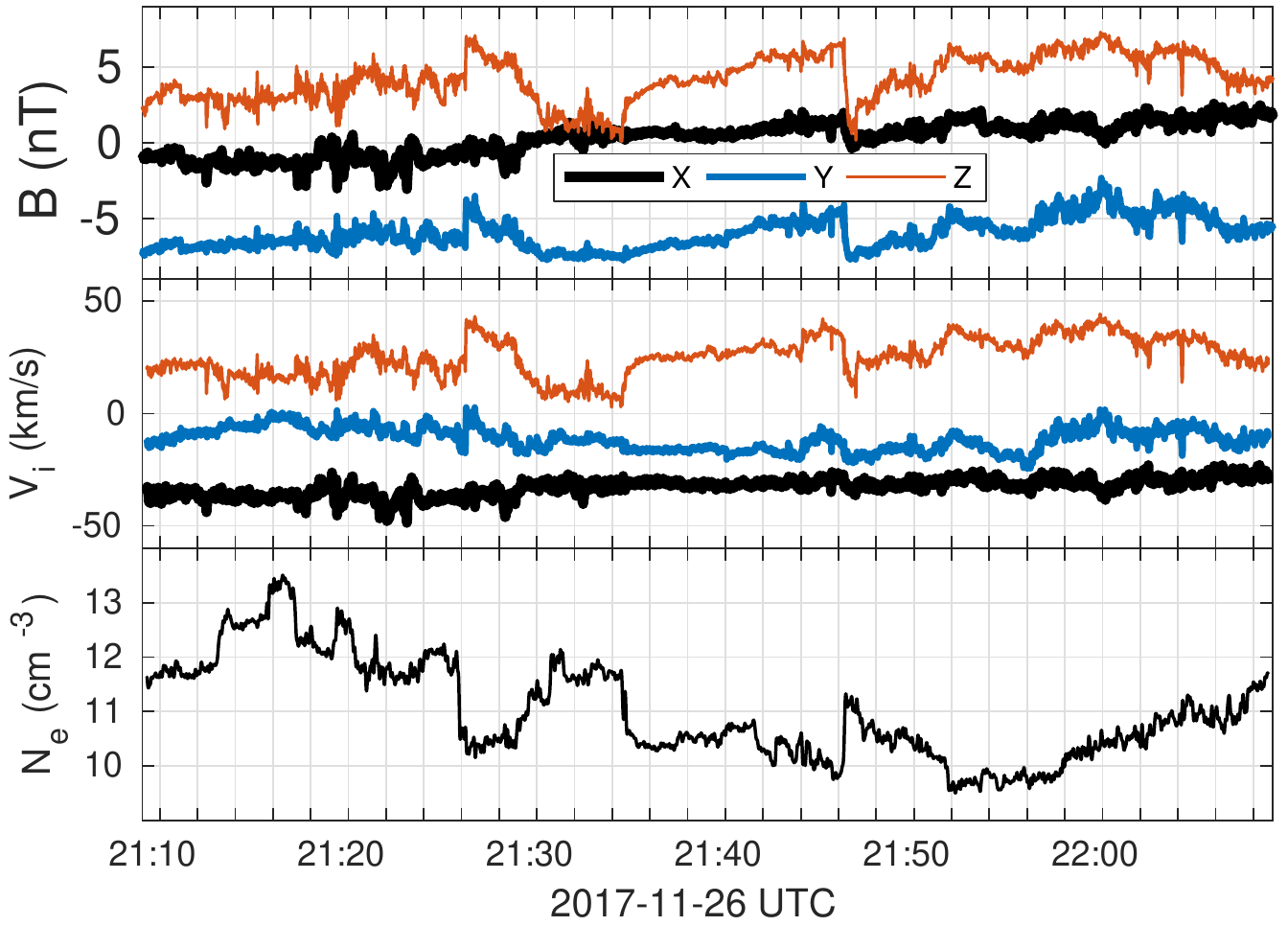}
		\caption{Time series plot of magnetic field (top panel), ion velocity (middle panel), and electron density (bottom panel) in GSE coordinate for the solar wind interval studied in the main text. The X component of velocity has been shifted by 300 km/s.}
		\label{fig:sw}
	\end{center}
\end{figure}

Figure~\ref{fig:sw} shows the time series of the FGM-measured magnetic field, the FPI/DIS measured ion velocity, and the FPI/DES measured electron density for the selected solar wind interval. For clarity in presentation, we have shifted the X component of velocity  by 300 km/s.
Due to limitations 
of FPI measurements in the solar wind, some systematic uncertainties 
exist in the higher-order plasma moments such as 
temperature. Therefore, we cross-check the average parameters of the selected solar wind interval (see main article) with \textit{Wind} data \cite{Ogilvie1995SSR, Lepping1995SSR}. The average density, velocity, and magnetic field values are in good agreement. However, significant discrepancy is found in the temperature and consequently the proton beta values between the FPI instrument and \textit{Wind} data. The FPI estimates a proton beta value of $0.8$ but the \textit{Wind} measurements estimate $\beta_{\mathrm{i}} = 0.4$. {Temperatures are likely overestimated due to the relatively large energy-angle bandwidths of the FPI instrument~\cite{Pollock2016SSR} compared to \textit{Wind} measurements. Given these known limitations~\citep[see][]{Bandyopadhyay2018aApJ} of the FPI instruments in the solar wind, we use the \textit{Wind} measurements of temperature and proton $\beta$ for this study.}

\begin{figure}
	\begin{center}
		\includegraphics[width=\linewidth]{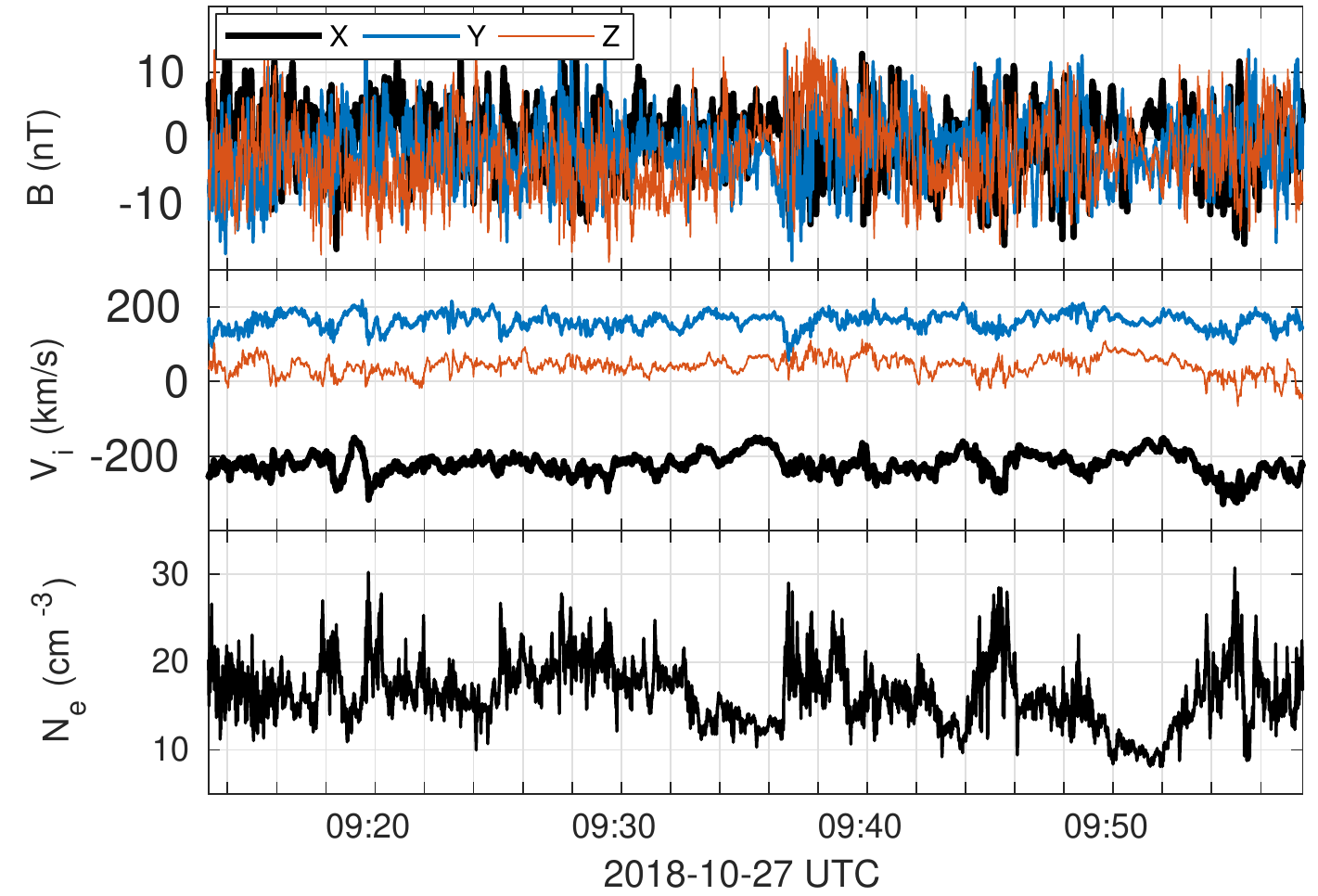}
		\caption{Time series plot of magnetic field (top panel), ion velocity (middle panel), and electron density (bottom panel) in GSE coordinate for the magnetosheath interval studied in the main text.}
		\label{fig:msh}
	\end{center}
\end{figure}

We plot the the magnetic field, the ion velocity, and the electron density for the chosen magnetosheath sample in figure~\ref{fig:msh}.

We recall that the generalized Hall-MHD third-order law in three-dimension is given by~\cite{Galtier2008PRE,Hellinger2018ApJL,Ferrand2019}
\begin{equation}
Y + \frac{1}{2}H=-\frac{4}{3}  \varepsilon \ell
\label{hall_supp},
\end{equation}
and in two dimensional system, like the simulations presented here, the $3$ in the denominator is replace by $2$.

\begin{figure}
	\begin{center}
		\includegraphics[width=\linewidth]{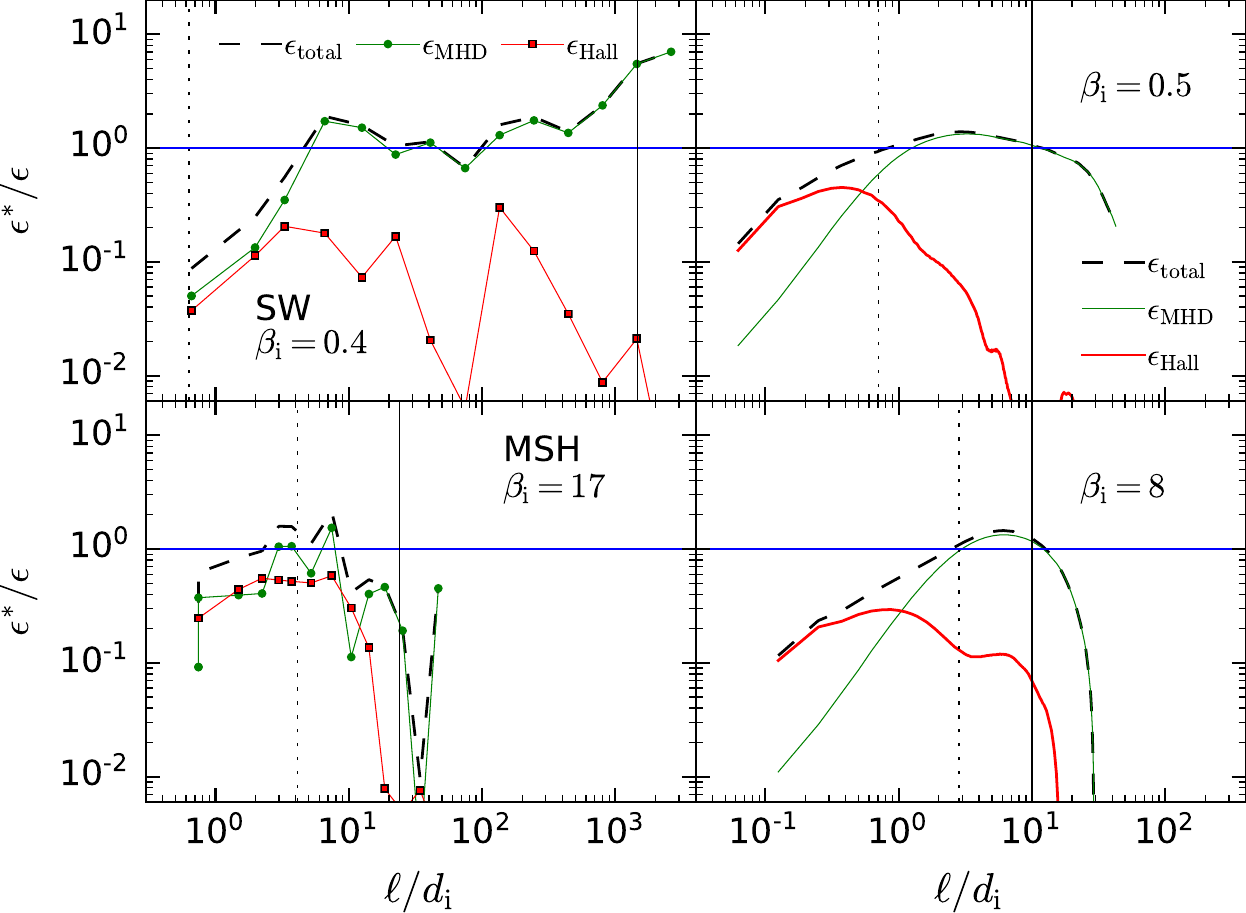}
		\caption{Energy cascade rate derived from the generalized Hall-MHD third-order law, normalized to the global decay rate.}
		\label{fig:eps}
	\end{center}
\end{figure}

For a more quantitative comparison between the simulation and MMS observation results, some type of normalization to the energy cascade rate, extracted from the Hall-MHD generalized third-order law, is in order. A natural choice of normalization is to compensate the inertial-range energy flux with some form of the global decay rate. If the inertial-range transfer rate, derived from equation~\ref{hall_supp}, yields the average energy loss rate in the system, the normalized value is expected to be close to unity.

In hybrid-kinetic simulations, the energy decay rate can be calculated \emph{exactly}, from the resistive heating:
\begin{eqnarray}
\epsilon = - \frac{\partial E}{\partial t} = \eta \langle \mathbf{\nabla} \mathbf{b} \mathbf{:} \mathbf{\nabla} \mathbf{b} \rangle \label{eq:rsist},
\end{eqnarray}
where, $\eta$ is the resistivity. The same cannot be done for the MMS observations, since the  resistivity in weakly-collisional plasma are not defined. However, a straightforward application of a von K\'arm\'an decay phenomenology, generalized to MHD ~\cite{Hossain1995PoF,Wan2012JFM}, gives a reasonable estimate of the global energy decay rate, at the energy-containing scales.
\begin{eqnarray}
\epsilon^{\pm} = - \frac{\partial (Z^{\pm})^2}{\partial t} = \alpha_{\pm} \frac{(Z^{\pm})^2 Z^{\mp}}{L_{\pm}}\label{eq:epspm},
\end{eqnarray}
where $\alpha_{\pm}$ are positive constants and $Z^{\pm}$ are the rms
fluctuation values of the 
Elsasser variables defined as $\mathbf{z}^{\pm}(t) = \mathbf{v}(t) \pm \mathbf{b}(t)$. The total energy decay rate can be calculated from the decay rate of the ``Elsasser energies", $\epsilon^{\pm}$, as
\begin{eqnarray}
\epsilon = \frac{(\epsilon^{+} + \epsilon^{-})}{2} \label{eq:eqpsvk}.
\end{eqnarray}
Figure~\ref{fig:eps} shows the MMS observation and simulation energy cascade rates, normalized in this manner. The left two panels are from MMS data and here, the different kinds of energy fluxes $(\epsilon^{*})$ are normalized by the von K\'arm\'an estimate of decay equation~$(\ref{eq:eqpsvk})$. A roughly flat scaling is observed in the inertial range for both intervals, implying a scale-invariant energy transfer in these length scales. 

The right two panels in figure~\ref{fig:eps} plot the results obtained from the simulation data~\cite{Hellinger2018ApJL,Franci2016ApJ}, but now normalized to the resistive heating rate, obtained from Eq.~\ref{eq:rsist}. Although the horizontal scaling is poorly defined here, compared to the observations, the values of the normalized energy flux are close to the MMS results.

{\section{Parameter Study of the Hybrid Simulations: Effect of  System Size}\label{sec:par}
	\begin{figure}
		\begin{center}
			\includegraphics[width=0.9\linewidth]{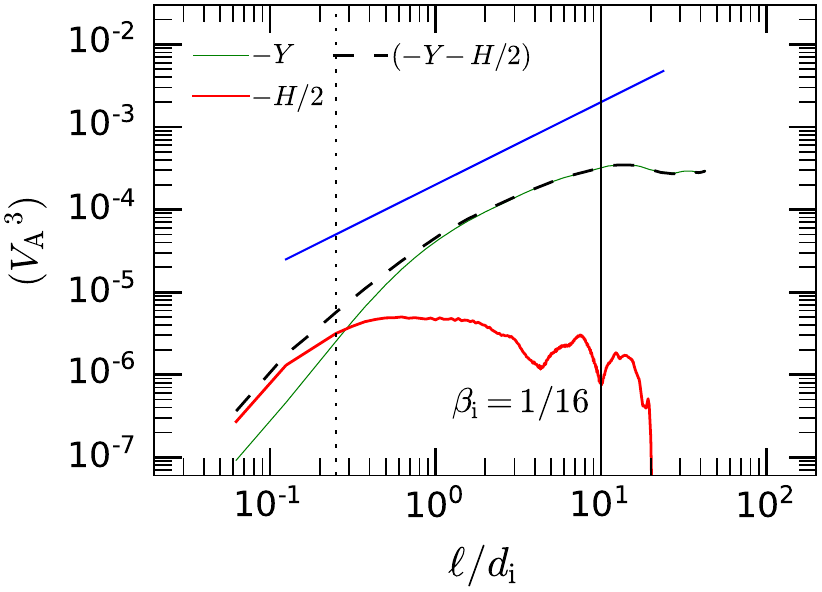}
			\includegraphics[width=0.9\linewidth]{fig_YHr_yagB.pdf}	\includegraphics[width=0.9\linewidth]{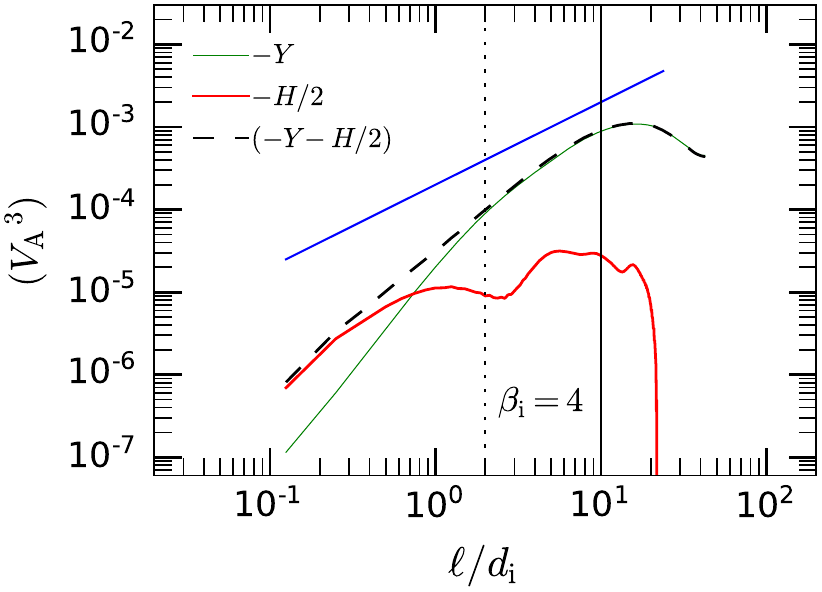}
			\includegraphics[width=0.9\linewidth]{fig_YHr_yagRb8.pdf}
			\caption{Variation of the Hall MHD third-order scaling with different scale separation $(L_{\mathrm{corr}}/\rho_{\mathrm{i}})$.}
			\label{fig:para}
		\end{center}
	\end{figure}
	As mentioned in the main text, a major weakness of the hybrid simulations is that those are necessarily of low Reynolds number. The Reynolds number can be conveniently gauged by the degree of scale separation, i.e., how far apart the kinetic scales (e.g., ion inertial length, ion gyro radius) lie from the energy-containing scales (which can be estimated by the correlation length). To examine that effect, we show here the Yaglom law scaling (Eq.~\ref{hall_supp}) for a number of hybrid simulation runs with varying $L_{\mathrm{corr}} / \rho_{\mathrm{i}}$. The scale separation decreases gradually from top to the bottom most panel in Fig.~\ref{fig:para}. It is clear that as the scale separation decreases the linear scaling of the total third-order structure function, $-(Y + H/2)$, becomes weaker. A similar study was carried out by Parashar \textit{et al.}~\cite{Parashar:ApJL2015}, using the onset of the von K\'arm\'an decay as a diagnostic. Parashar \textit{et al.}~\cite{Parashar:ApJL2015} found that kinetic plasmas start to exhibit fluid-like behavior at large-scale, when the energy-containing scales are separated from the kinetic scales by a factor of about 100.}

{\section{Additional MMS intervals}\label{sec:add}
	In Fig.~\ref{fig:add} we show the Yaglom law scaling (Eq.~\ref{hall_supp}) for a few additional MMS intervals in the magnetosheath with a broad range of proton beta. These intervals are also used in the original paper to obtain the variation of $2Y/H$ with $\beta_{\mathrm{i}}$.
	\begin{figure}[hb!]
		\begin{center}
			\includegraphics[width=\linewidth]{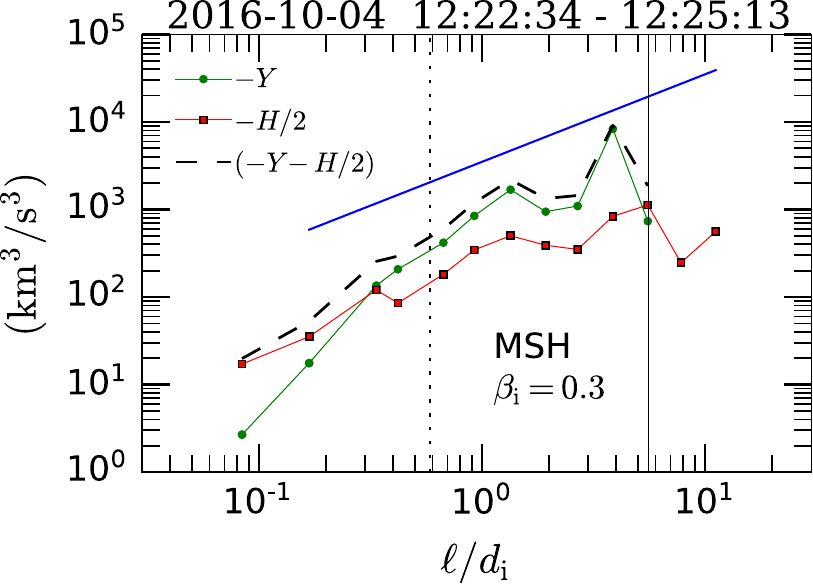}
			\includegraphics[width=\linewidth]{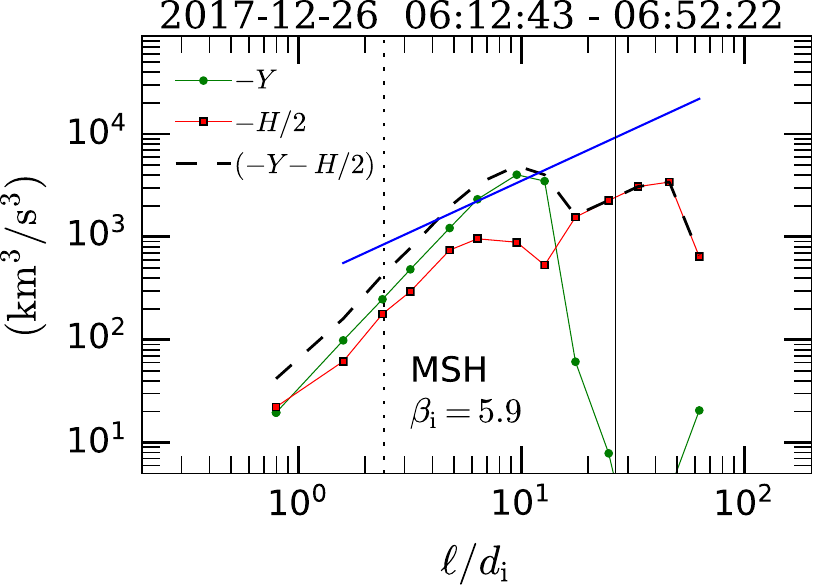}
			\includegraphics[width=\linewidth]{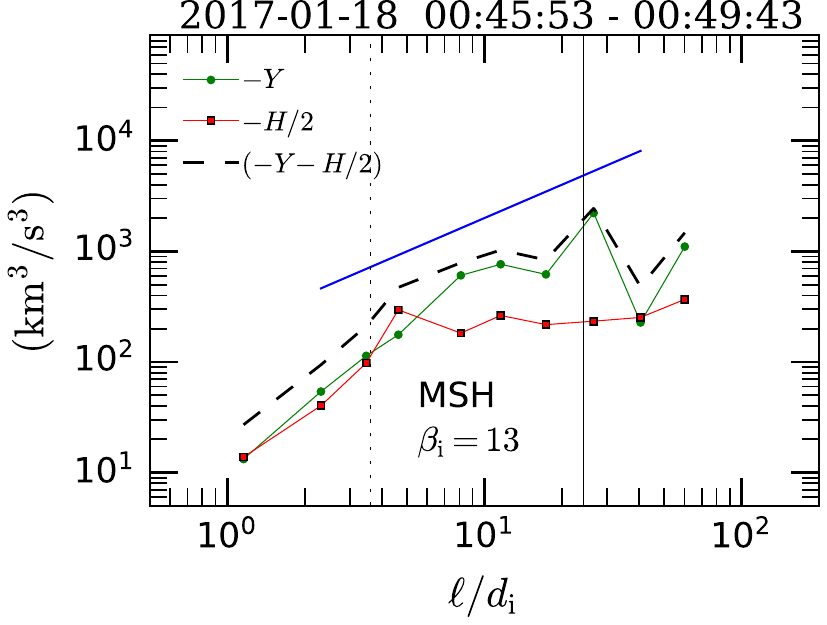}
			\caption{Hall-MHD Yaglom law scaling for additional MMS intervals.}
			\label{fig:add}
		\end{center}
	\end{figure}}

\bibliographystyle{apsrev4-1}\texttt{}
%

\end{document}